\documentclass{article}
\usepackage{hiph-preprint}
\usepackage{graphicx}
\volnumber{22} \issuenumber{1} \edyear{2005}                             
\frompage{000} \topage{000}                                              
\recrevdate{1 January 2005}                                              
\def\rightmark{Di-jet Shape Modification in Heavy Ion Collisions}
\def\leftmark{P. Constantin}

\title{Di-jet Shape Modification in Heavy Ion Collisions}
\authors{
{Paul Constantin (for the PHENIX Collaboration) %
\index{Constantin, P.}
}\\[2.812mm]
{\normalsize \hspace*{-8pt} Los Alamos National Laboratory,
Los Alamos, NM 87545, USA\\[0.2ex]
}}

\abstract{We present preliminary results from intermediate $p_T$
(1-5$GeV/c$) di-hadron azimuthal correlations induced by hadronic
di-jets produced in $AuAu$ collisions at $\sqrt{s_{NN}} = 200 GeV$.
The near-side ($\Delta\phi\sim0$) has a typical single-peaked
structure which broadens with the centrality of the collision.
A qualitatively new phenomenon shows up in the shape of the
away-side ($\Delta\phi\sim\pi$): it has a symmetric, double-peaked
structure in central and mid-central collisions.}


\makeindex
\begin{document}

\maketitle

\section{Introduction}\label{intro}

    The $AuAu$ data collected during the 2002 RHIC run established
several new exciting results in the area of hadronic jet
modification by a hot and dense QCD medium: leading hadron
suppression \cite{bib1}, away-side disappearance at high $p_T$
\cite{bib2}, and the strong (non-Gaussian) modification of the
away-side at intermediate $p_T$ \cite{bib3}.

    The goal of the much higher statistics 2004 RHIC run is a
systematic study of these effects. This paper addresses the
particular issue of how the di-jet shape is modified by
interacting with the medium formed in $AuAu$ collisions at high
energy.

\section{Di-jet Shapes from Di-hadron Azimuthal
Correlations}\label{method}

    In order to correct for the limited acceptance of the PHENIX detector,
we build our statistical di-hadron azimuthal correlations by
dividing with mixed event azimuthal distributions
\cite{bib3,bib4}.

\begin{figure}[htb]
\begin{center}
\resizebox{6.5 cm}{4.5 cm}{\includegraphics{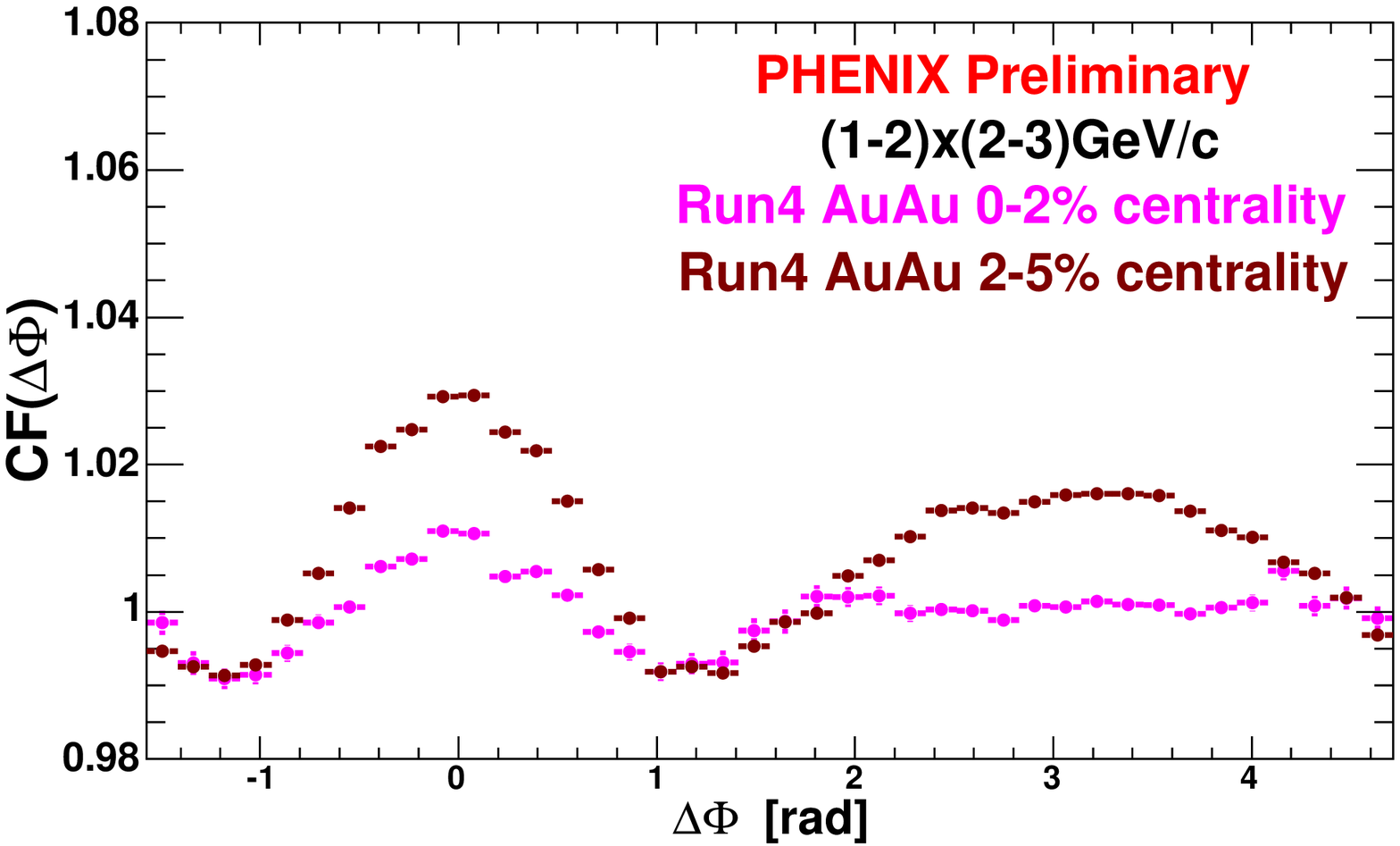}}
\resizebox{6.0 cm}{4.5 cm}{\includegraphics{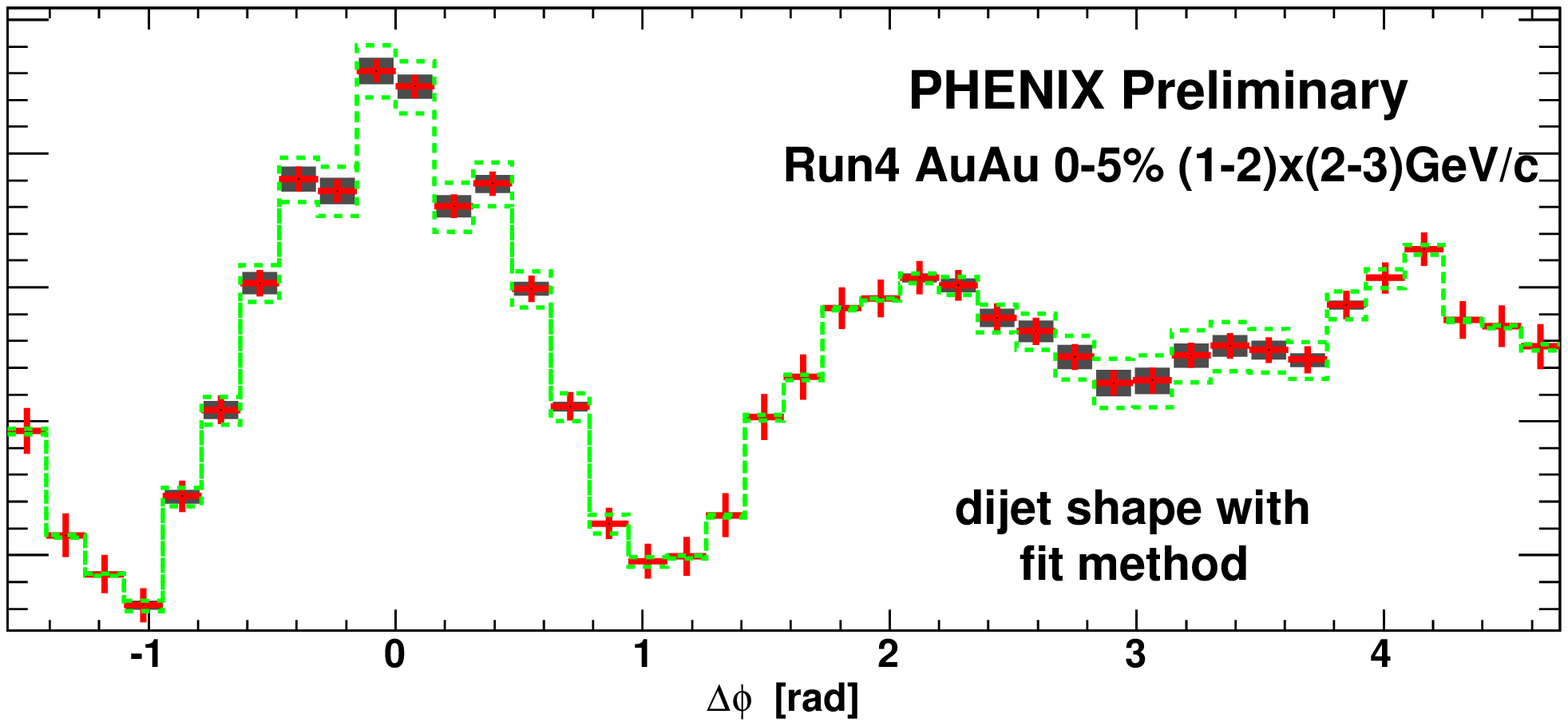}}
\caption[]{Left: di-hadron azimuthal correlation functions in very
central (pink - 0-2\% and brown - 2-5\%) $AuAu$ collisions,
normalized to match at $\Delta\phi = 1rad$. Right: the extracted
shape of the di-jet azimuthal correlation function; black boxes
and green dashed lines represent $1\sigma$ and $2\sigma$
systematic errors.} \label{CorrelFig}
\end{center}
\end{figure}

    An example of such azimuthal correlations is shown in the left
panel of Figure \ref{CorrelFig} for (2-3)GeV/c trigger hadrons and
(1-2)GeV/c associated hadrons in the 0-2\% and 2-5\% most central
$AuAu$ collisions at $\sqrt{s_{NN}} = 200 GeV$. A striking feature
is the wide flat plateau around $\Delta\phi\sim\pi$, before
background subtraction. Considering the $\cos(2\Delta\phi)$ shape
of the background (see Equation (\ref{CFeq})), it is obvious that
the resulting di-jet shape will have a local minimum at
$\Delta\phi\sim\pi$ after subtraction. This can be seen indeed in
the extracted di-jet induced correlation function shown in the
right panel of the same figure.

    Two sources contribute to angular correlations in this $p_T$
range - a global correlation of all hadrons with the collision
reaction plane \cite{bib6}, and the di-jet induced correlation
$J(\Delta\phi)$:

\begin{equation}
    C(\Delta\phi) = \xi (1+2v_2(p_T^{trigg})v_2(p_T^{assoc})\cos(2\Delta\phi)) + J(\Delta\phi)
\label{CFeq}
\end{equation}

    The presence of a momentum conservation source proportional to
$-\cos(\Delta\phi)$, observed at much lower associated hadron
momenta \cite{bib9}, is under study.

    We use the measured $v_2$ of charged hadrons in the PHENIX central arms
($|\eta|<0.35$) with respect to the reaction plane orientation in
the BBCs (Beam-Beam Counters with $3<|\eta|<3.9$) from
\cite{bib6}.

    We then fit the correlation functions with the right hand side
of Equation (\ref{CFeq}), where the di-jet term is parameterized
as a Gaussian for the near side and a symmetric double Gaussian
for the away side, hence $J(\Delta\phi)$ has the functional form:

\begin{equation}
A_N \exp\Big(\frac{-\Delta\phi^2}{2w_N^2}\Big) + A_A
\Big(\exp\big(\frac{-(\Delta\phi-\pi-D)^2}{2w_A^2}\big) +
     \exp\big(\frac{-(\Delta\phi-\pi+D)^2}{2w_A^2}\big)\Big)
\label{FITeq}
\end{equation}

    with six free parameters: the background level $\xi$, the
near/away di-jet amplitudes $A_{N,A}$, and the di-jet shape
parameters $w_N$ (near width), $w_A$ (away width), and $D$ (away
splitting). This functional form provides an excellent description
of the di-jet correlation functions for a wide selection of
centralities and momenta.

    The main source of systematic errors comes from the background
estimation, with its two components: the $v_2$ error, which is
propagated from the published measurement \cite{bib6}, and the
$\xi$ error. The latter is estimated by using an independent
method, called the ZYAM (Zero di-jet Yield At Minimum) method
\cite{bib3,bib5}.

\section{Di-jet Shape Parameters at Intermediate $p_T$}\label{main}

    Using the above method, we extract the di-jet induced correlation
functions for $AuAu$ and $dAu$ collisions at $\sqrt{s_{NN}} =
200GeV$ in various centrality classes and $p_T$ regions. The
qualitatively new feature, shown in the right panel of Figure
\ref{CorrelFig} and present only for di-jet induced correlation
functions in relativistic heavy ion collisions, is the splitting
of the away side into a symmetric, double-peaked structure.

\begin{figure}[htb]
\begin{center}
\resizebox{9.0 cm}{7.0 cm}{\includegraphics{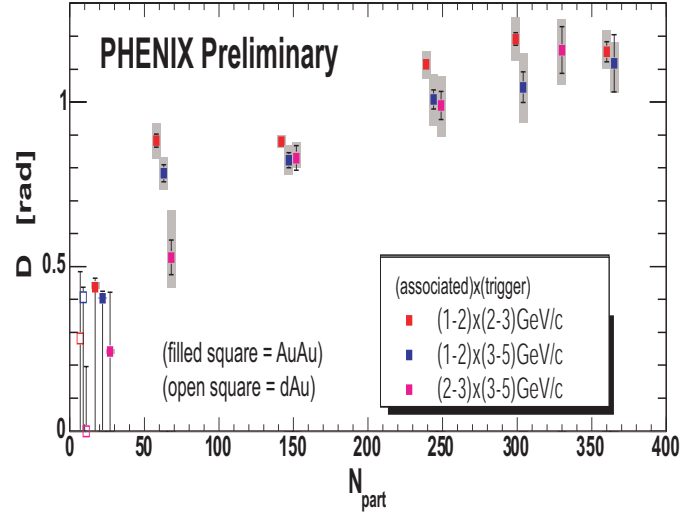}}
\caption[]{Centrality and momentum dependence of the splitting
parameter $D$.} \label{SplitFig}
\end{center}
\end{figure}

    The position of the away side peaks is parameterized by $D$.
Its centrality and momentum dependence are shown in Figure
\ref{SplitFig}: $D$ rapidly increases in mid-peripheral collisions
and then has a slower increase towards more central collisions. It
also slowly decreases with both the trigger and associated hadron
momenta.

    Several theoretical models to explain this result have been proposed
already. In one of them, partons with velocities larger than the
speed of sound in a liquid QGP produce shock waves propagating in
a Mach cone with respect to parton's momentum; $D$ is then the
Mach angle, hence it measures the QGP speed of sound \cite{bib7}.
In another model, partons with velocities larger than the speed of
light in a QGP with bound partonic states produce gluon radiation
in a Cherenkov cone with respect to parton's momentum; $D$ is then
the Cherenkov angle, hence it measures the QGP index of refraction
\cite{bib8}.

\begin{figure}[htb]
\begin{center}
\resizebox{6.0 cm}{6.0 cm}{\includegraphics{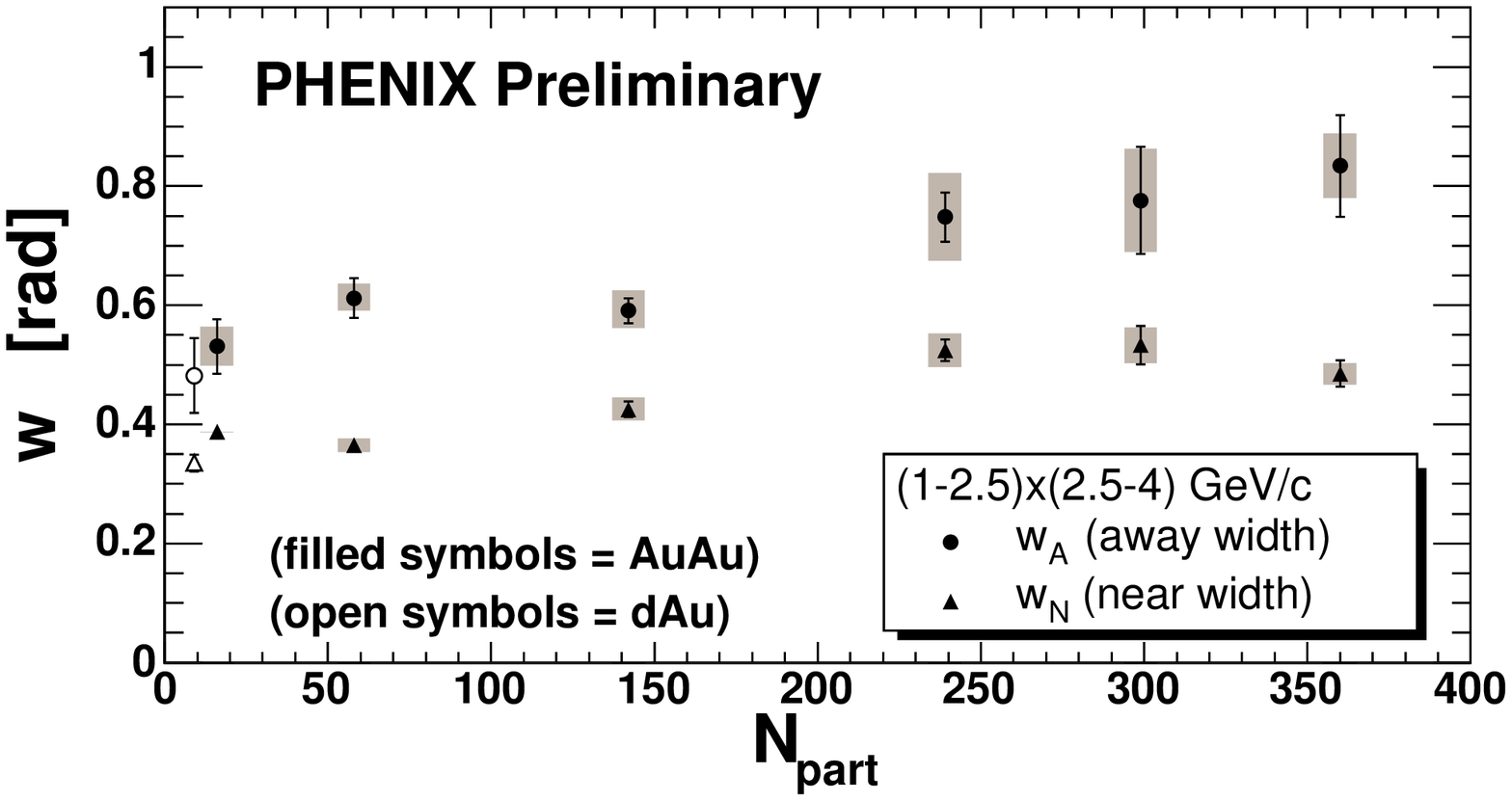}}
\resizebox{6.0 cm}{6.0 cm}{\includegraphics{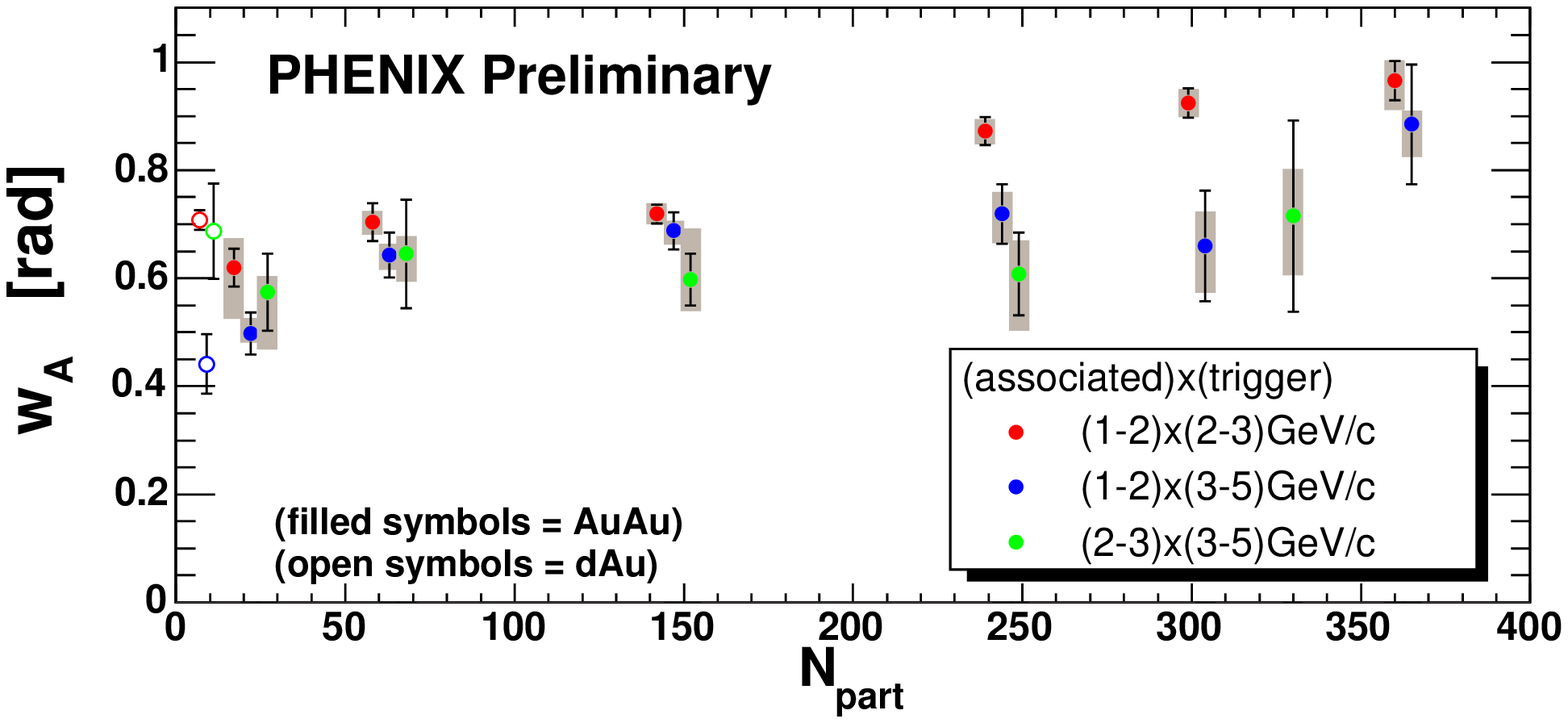}}
\caption[]{Left: centrality dependence of the near width $w_N$
(triangles) and of the away width $w_A$ (circles). Right: momentum
dependence of the away width $w_A$.} \label{WidthFig}
\end{center}
\end{figure}

    The left panel of Figure \ref{WidthFig} shows the centrality
dependence of the near side width $w_N$ (triangles) and of the
away side width $w_A$ (circles). Both widths broaden with
collision centrality at intermediate $p_T$. The right panel of
this figure shows the momentum dependence of the away width $w_A$.
A trend from broadening at intermediate $p_T$ towards centrality
independence at higher trigger and associated momenta can be
observed, but higher statistics is needed.

\vfill\eject

\begin{thebibliography}{99}

\bibitem{bib1} S.S. Adler et al, {\it Phys. Rev. Lett.} {\bf 91} (2003) 072301

\bibitem{bib2} C. Adler et al, {\it Phys. Rev. Lett.} {\bf 90} (2003) 082302

\bibitem{bib9} C. Adler et al, {\it Phys.Rev.Lett.} {\bf 95} (2005) 152301

\bibitem{bib3} S.S. Adler et al, {\it nucl-ex/0507004}

\bibitem{bib4} N. Grau, {\it nucl-ex/0511046}

\bibitem{bib5} J. Jia, {\it Acta Phys. Hung.} {\bf A22} (2005), {\it nucl-ex/0510060}

\bibitem{bib6} S.S. Adler, et al, {\it Phys.Rev.Lett.} {\bf 91} (2003) 182301

\bibitem{bib7} J. Casalderrey, E. Shuryak and D. Teaney, {\it hep-ph/0411315}\\
               L.M. Saratov, H. Stocker and I.N. Mishustin, {\it Phys.
               Lett.} {\bf B 627} (2005) 64 \\
               T. Renk and J. Ruppert, {\it hep-ph/0509036}

\bibitem{bib8} V. Koch, A. Majmuder and X.-N. Wang, {\it nucl-th/0507063}

\end{thebibliography}
\end{document}